\documentclass[twoside,english]{elsarticle}
\usepackage[T1]{fontenc}
\usepackage[latin9]{inputenc}
\usepackage{verbatim}
\usepackage{prettyref}
\usepackage{amsmath}
\usepackage{amssymb}
\usepackage{graphicx}
\usepackage{wasysym}

\makeatletter
\journal{Example: Annals of Physics}

\makeatother

\usepackage{babel}
\begin{document}

\begin{frontmatter}{}

\title{Multi-valued vortex solutions to the Schrödinger equation and radiation\tnoteref{t1,t2}}

\author[rvt]{Mark Davidson\fnref{}}

\ead{mdavid@spectelresearch.com}

\address[rvt]{Spectel Research Corp., Palo Alto, CA USA}
\begin{abstract}
This paper addresses the single-valued requirement for quantum wave
functions when they are analytically continued in the spatial coordinates.
This is particularly relevant for de Broglie-Bohm, hydrodynamic, or
stochastic models of quantum mechanics where the physical basis for
single-valuedness has been questioned. It first constructs a large
class of multi-valued wave functions based on knotted vortex filaments
familiar in fluid mechanics, and then it argues that for free particles,
these systems will likely radiate electromagnetic radiation if they
are charged or have multipolar moments, and only if they are single-valued
will they definitely be radiationless. Thus, it is proposed that electromagnetic
radiation is possibly the mechanism that causes quantum wave functions
to relax to states of single-valuedness, and that multi-valued states
might possibly exist in nature for transient periods of time. If true,
this would be a modification to the standard quantum mechanical formalism.
A prediction is made that electrons in vector Aharonov-Bohm experiments
should radiate energy at a rate dependent on the solenoid's magnetic
flux.
\end{abstract}

\end{frontmatter}{}

\section{Introduction}

The question whether the Schrödinger equation must always be single-valued
occupied a number of the founders of early quantum theory, starting
with Schrödinger himself \citep{schrodingerthemultivaluedness}, then
notably Pauli \citep{pauli1939uberein}, and subsequently a number
of others \citep{riess1973singlevalued,merzbacher1962singlevaluedness,yang1984gaugefields}.
The question received renewed interest after the Aharonov-Bohm effect
was discovered \citep{aharonov1959significance}. The general consensus
reached was that even in the Aharonov-Bohm systems the wave function
must still be single-valued \citep{peshkin1989theaharonovbohm,yang1984gaugefields,merzbacher1962singlevaluedness},
but there were some dissenters \citep{siverman1985onthe}. The subject
received a new burst of interest from the insights of Wallstrom regarding
the non-obvious motivation for the single-valued constraint in de
Broglie-Bohm, hydrodynamic, and stochastic formulations of quantum
mechanics \citep{wallstrom1989onthe,wallstrom1990afiniteenergy,wallstrom1990thestochastic,wallstrom1994onthe,wallstrom1994inequivalence}.
Goldstein also independently commented on multi-valued wave functions
in stochastic mechanics \citep{goldstein1987stochastic}. Derakhshani
has recently suggested an interesting explanation for the single-valued
constraint based on zitterbewegung \citep{derakhshani2015asuggested,derakhshani2016asuggested}.
Smolin has also considered this issue \citep{smolin2006couldquantum}.
Here I take a different approach to this question which is based on
electromagnetic radiation. 

The prototypical example of a multi-valued wave function is one of
the form $\psi(x)=f(x)e^{il_{z}\varphi}$, where $f(x)$ is a single-valued
function and where $\varphi$ is the azimuthal angle, and $l_{z}$
is a real constant. The wave function is single-valued only if $l_{z}=\pm n,\:n\in\mathbb{Z}$.
The single-valued constraint leads to quantization effects. In this
simple case, when $l_{z}$ is not an integer, we can consider the
wavefunction to have multiple Riemann sheets that differ from one
another by constant phase factors of the form $e^{il_{z}2\pi m}$
for $m$ integer. The fact that the different sheets differ by a constant
phase factor which is independent of x, is important, and consequently
I shall consider only multi-valued wave functions which have this
property here.

Firstly I introduce a broad class of suitable multi-valued wave functions.
I do this by borrowing results from the theory of vortex filaments
in fluid mechanics. The fluid version of the Biot-Savart law allows
one to construct a local potential function which is generated by
a vortex filament taking an arbitrary stringlike shape, either closed
or open . This includes arbitrary knots and links, and this function
is generally multi-valued when analytically continued in $\mathbf{x}$.
From these known solutions, I construct multi-valued solutions of
the Schrödinger equation. For the usual azimuthal example, the string
would be the whole z axis, and be infinitely long. The analysis presented
applies to such cases as well as to knots and links. The subject of
quantum vortices applied to the Schrödinger equation is not new \citep{berry2001knotting,bialynicki-birula2000motionof,lloyd2017electron,bliokh2017theoryand}.
Here we generalize these treatments to multi-valued wave functions.

Then I show, by using a nonlinear identity of the Schrödinger equation,
that the multi-valued linear Schrödinger equation can be replaced
by an equivalent single-valued but nonlinear equation. I then argue
from this that if the particle is charged, then the nonlinear term
will likely produce spontaneous bremsstrahlung even for a free particle,
so that the multi-valued solution may not be stable to radiative decay.
Consequently, I argue that the equilibrium state of the system that
is approached must be single-valued, and since this could account
for the observed fact of single-valuedness, at least for charged particles,
that perhaps the single-valued condition of quantum mechanics might
not be universally true, and in particular it might be violated for
short time periods where, after a collapse of the wave function for
example, a transient multi-valued wave function is produced which
then subsequently decays into a single-valued one. Although this radiative
relaxation to single-valuedness is easiest to understand for charged
particles, it might be expected to hold for neutral particles too
if they have some multipolar electromagnetic moments, since these
too would radiate when accelerated. Neutrons or neutral atoms are
examples. So we might expect that multi-valued wave functions if ever
created would typically be short-lived for them as well. The exceptions
might be neutrinos, or dark matter particles (if they exist), where
the lack of any electromagnetic interaction would allow multi-valued
wave functions to persist for longer periods of time. This transient
multi-valuedness might yield experimentally detectable effects. We
discuss one such test below in the vector Aharonov-Bohm discussion.
Also, these results might also prove to be useful in the mathematical
theory of knots. 

When restricted by the single-valued constraint, the vortex solutions
here are similar to the quantum vortex solutions in superconductors,
and superfluids. Here they are considered as solutions to the single
particle Schrödinger equation with or without a potential. I'm not
aware of the general knot solutions found here having been considered
previously in the physics literature, although a number of special
cases were elegantly analyzed in \citep{bialynicki-birula2000motionof}.

\section{A multi-valued initial state based on a knotted vortex filament solution
using the Biot-Savart law}

It is well known that the single particle Schrödinger equation can
be cast in the form of inviscid fluid mechanical equations by the
Madelung transformation \citep{madelung1926eineanschauliche,madelung1927quantentheorie}.
The Euler equations take the following form in a conservative force
field $V$ with pressure $p$, and density $\rho$

\begin{equation}
\left(\frac{\partial}{\partial t}+\mathbf{u}\cdot\nabla\right)\mathbf{u}=-\frac{1}{\rho}\nabla p-\nabla V
\end{equation}

\noindent and with the Madelung ansatz for the quantum force, this
becomes

\begin{equation}
\left(\frac{\partial}{\partial t}+\mathbf{u}\cdot\nabla\right)\mathbf{u}=-\nabla\left(Q+V\right)
\end{equation}

\noindent where 

\begin{equation}
Q=-\frac{\hslash^{2}}{2m}\frac{\triangle\sqrt{\rho}}{\sqrt{\rho}}
\end{equation}

\noindent and where $\rho$ is a conserved density. Of course, these
equations are identical to those in Bohmian mechanics for the single
particle case \citep{bohm1952asuggested}. For multiple particle states,
especially when entangled, the Bohmian theory describes the system
more easily than a hydrodynamic picture as it does away with a universal
guiding fluid for all the particles, but if the presence of the particles
in the fluid can affect the fluid currents that the other particles
see, as in a theory in which the particles are actually solitons for
example, then it's still conceivable to construct a many-particle
hydrodynamic model of quantum mechanics for this case too. These equations
can also be considered as diffusion equations for a Brownian motion
process as in stochastic mechanics \citep{nelson1967dynamical}. In
fact Wallstrom's original interest in this subject arose from the
equations of stochastic mechanics \citep{wallstrom1989onthe}. 

\begin{figure}
\begin{centering}
\includegraphics[scale=0.25]{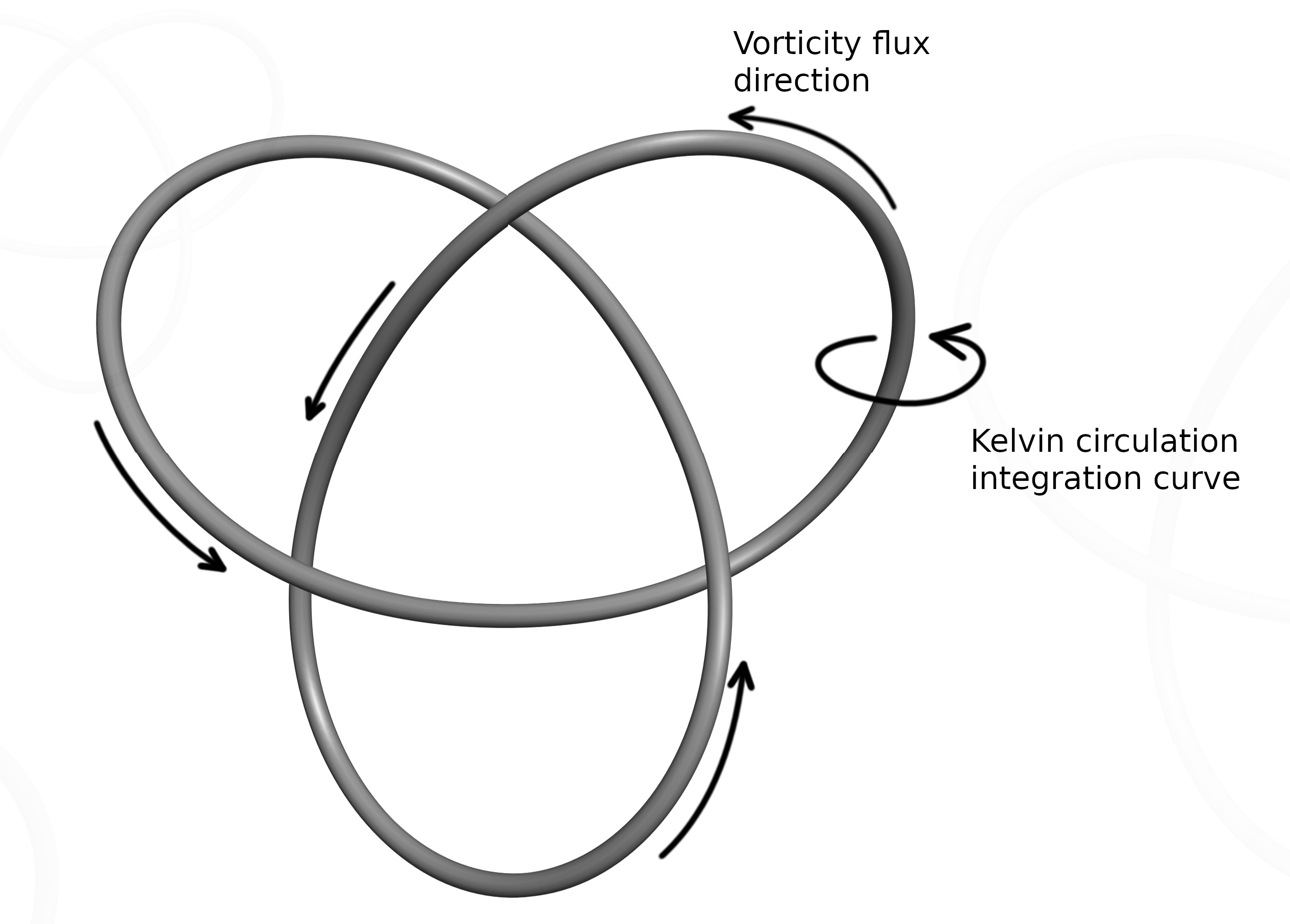}
\par\end{centering}
\caption{\label{fig:Vortex-filament-knot}Vortex filament knot}
\end{figure}

We first consider an incompressible fluid which satisfies $\nabla\cdot\mathbf{u}=0$
and is described by the knotted vortex filament solutions from 3D
fluid mechanics \citep{saffman1992vortexdynamics,kauffman1987onknots,kauffman2006knottheory}.
Let $\mathbf{u}(\mathbf{x})$ denote such a solution, which is written
in terms of the ``Biot-Savart'' law applied to an inviscid incompressible
fluid with a vorticity filament forming a knot, as in \prettyref{fig:Vortex-filament-knot}.
The vortex filament is a thin tube inside of which the vorticity is
non-zero, and pointing along the tangent to the tube. Outside of the
filament the vorticity is zero.

A 3-space curve for the shape of a filament knot is given by a vector
function $\mathbf{R_{f}}(\sigma)$ where $\sigma$ is arc-length along
the curve which is assumed to be continuous and smooth, and the tangent
vector $d\mathbf{R_{f}}/d\sigma$ is in the direction given by the
right hand rule applied to the circulation about the filament. For
example, the simple trefoil knot illustrated in \prettyref{fig:Vortex-filament-knot}
is topologically equivalent to the following space curve, where $\beta$
varies from $0$ to $2\pi$

\begin{equation}
\begin{array}{c}
x=sin(\beta/2\pi)+2*sin(2*\beta/2\pi)\\
y=cos(\beta/2\pi)-2*cos(2*\beta/2\pi)\\
z=-sin(3*\beta/2\pi)
\end{array}\label{eq:Trefoil knot equation}
\end{equation}

In this simple example formula, $\beta$ is not equal to the arclength,
but one could reparametrize it in principle to express the curve in
terms of arclength $\sigma$ if needed. It is straightforward to also
include cases where the filament is not closed, but goes off to infinity
in both directions, or where there is more than one knot superimposed
with the same value of the vortex circulation constant forming a link.
The velocity field of the fluid at an arbitrary point $\mathbf{x}$
produced by a single closed filament is given by the ``Biot-Savart''
law with $\varGamma$ being the circulation constant of the vortex
as in section 2.3 of \citep{saffman1992vortexdynamics}:

\begin{equation}
\mathbf{u}_{\mathbf{f}}(\mathbf{x})=\frac{\varGamma}{4\pi}\varoint\frac{d\overrightarrow{\mathbf{\mathbf{\sigma}}}\times\left(\mathbf{x}-\mathbf{R_{f}}(\sigma)\right)}{\left|\mathbf{x}-\mathbf{R_{f}}(\sigma)\right|^{3}}=\frac{\varGamma}{4\pi}\nabla\times\varoint\frac{d\overrightarrow{\mathbf{\mathbf{\sigma}}}}{\left|\mathbf{x}-\mathbf{R_{f}}(\sigma)\right|}
\end{equation}

\noindent If we have more than one filament, we just add their $\mathbf{u}_{\mathbf{f}}$
together to get the resultant. As $\mathbf{u_{f}}(\mathbf{x})$ is
irrotational, it follows that in a simply connected domain $\varOmega_{x}$
which does not intersect the knot curve we have 

\begin{equation}
\mathbf{u_{f}}(\mathbf{\mathrm{x}})=\nabla\phi_{f}(x),\:\mathbf{x}\in\varOmega_{x}
\end{equation}

\noindent for some scalar function $\phi_{f}(x)$. This domain can
be expanded to include the whole space, but as it must exclude the
filament, the resulting domain then becomes multiply connected, and
consequently $\phi_{f}(x)$ will generally be multi-valued as $x$
is analytically continued around the filament as shown in figure 1.
Since $\nabla\cdot\mathbf{u_{f}}=0$ is required for an incompressible
fluid, it follows that

\begin{equation}
\triangle\phi_{f}(x)=0
\end{equation}

\noindent and it also follows from Stoke's theorem that

\begin{equation}
\varoint_{C}\mathbf{u}(\mathrm{\mathbf{x}})\cdot d\mathbf{x}=\varGamma W(C)
\end{equation}

\noindent where $W(C)$ is the winding or circulation number of the
integration loop around the filament. It follows that although $\phi_{f}(x)$
is multi-valued, the different values at the same $\mathbf{x}$ differ
by additive constants which do not depend on $x$. Consequently $\mathbf{u}(\mathbf{x})$,
being a gradient of $\phi_{f}$, is not changed by these constants,
and is single-valued in this case. Now if we add to $\mathbf{u}$
a second solenoidal velocity field $\mathbf{\mathrm{\mathbf{w}(x)}}$
which is derived from a single-valued potential, $\mathbf{w}(x)=\nabla\phi_{w}(x)$
which is valid for all $x$, then since $\varoint_{C}\mathbf{w}(x(\mathbf{s}))\cdot d\mathbf{s}=0$
for any closed curve $C$, we must then have that

\begin{equation}
\varoint_{C}\left(\mathbf{u}_{\mathbf{f}}(x(s))+\mathbf{w}(x(s))\right)\cdot d\mathbf{x}=\varGamma W(C)
\end{equation}

The combined velocity in this case no longer divergence free, and
it can describe a compressible fluid as is required for the initial
velocity field of a Schrödinger equation in the Madelung construction.
The vorticity filament will move advectively with the fluid as time
progresses, as described by the Kelvin circulation theorem.

\begin{figure}
\noindent \begin{centering}
\includegraphics[scale=0.7]{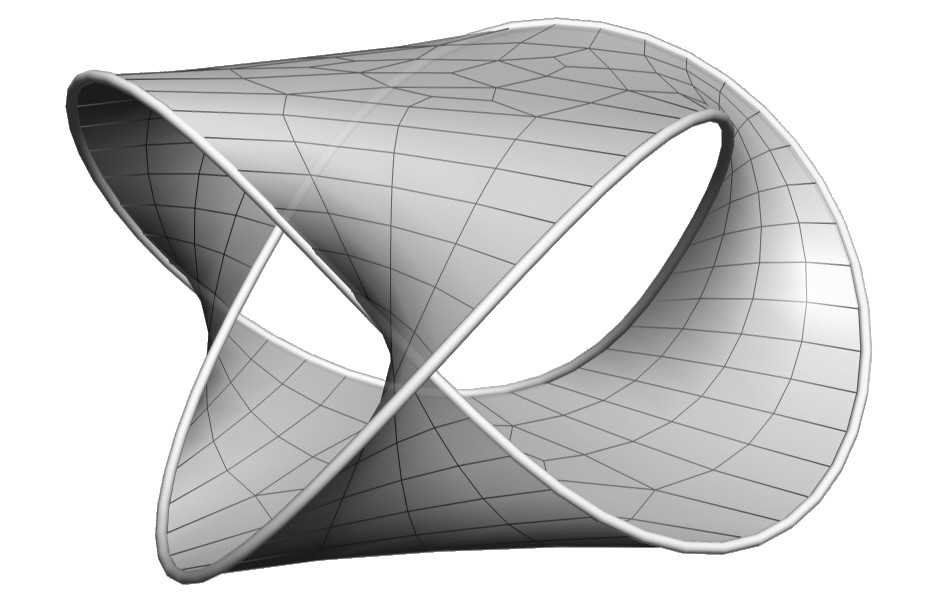}
\par\end{centering}
\caption{\label{fig:A-rendering-of-Seifert-surface}A rendering of a Seifert
surface for a Trefoil knot by SeifertView\citep{wijk2006visualization}}
\end{figure}

The task of finding the potential function $\phi_{f}(x)$ is identical
to two different problems in electromagnetism. The first and most
common is the magnetic field $\mathbf{B}$ generated by a charge current
$I$ flowing along a knot, and expressed as the gradient of a magnetic
scalar potential. The second is the problem of calculating the vector
potential $\mathbf{A}$ which is generated by a magnetic flux filament
pointing along a knot curve. The magnetic scalar potential is discussed
in many sources, for example \citep{jackson1999classical,franklin2017classical,portmann2018selfadjointness,portmann2019spectral,gross2004electromagnetic,kotiuga1987onmaking}.
A classical result for a vortex loop which is unknotted, ie. topologically
equivalent to a circle, is (equation 2.5 in \citep{saffman1992vortexdynamics})

\begin{equation}
\phi_{f}(x)=-\frac{\varGamma}{4\pi}\Omega(x)
\end{equation}
where $\Omega(x)$ is the solid angle subtended by the vortex loop
when viewed from the point $\mathbf{x}$. In integral form this is

\begin{equation}
\Omega(x)=\int_{S}\frac{\left(\mathbf{x}-\mathbf{R}\right)\cdot\mathbf{dS}}{\left|\mathbf{x}-\mathbf{R}\right|^{3}}
\end{equation}
where $R$ is a point on the surface, and the surface $S$ is bounded
by the loop. This result, originally due to Maxwell \citep{binysh2018maxwells},
can be applied to knotted vortex loops too by using a Seifert surface
for the knot \citep{seifert1935uberdas,wijk2006visualization}. These
are compact, connected, and oriented surfaces with the knot as its
boundary. A beautiful tool for visualizing them is a software program
called SeifertView, created by Professor Jarke van Wick, Technische
Universiteit Eindhoven \citep{wijk2006visualization}. An example
of a Seifert surface is shown in \prettyref{fig:A-rendering-of-Seifert-surface}.

The Seifert surface acts as a cut analogous to the Riemann sheet cut
familiar from complex analysis. If we analytically continue $\phi_{f}(x)$
along some curve, but avoid ever passing through the Seifert surface,
it will remain single-valued, but the difference between its value
on one side versus the opposite side of the surface will be non-zero,
and independent of where on the surface the discontinuity is calculated.
When dealing with links we can superimpose the potential functions
for the individual knots that make up the link because of the linearity
of Laplace's equation. The vorticity constant for the knots making
up a link need not be the same in general, but for our purposes here,
we shall assume that they are the same so that the discontinuity of
$\phi_{f}(x)$ across the surface will always be the same, up to a
sign. Links have Seifert surfaces too, and although they are not unique,
I presume that one always exists which can act as a suitable cut surface. 

\section{The Schrödinger equation in the de Broglie-Bohm-Madelung pilot wave
formalism}

We consider the single-particle Schrödinger equation

\begin{equation}
\left[-\frac{\hslash^{2}}{2m}\triangle+V\right]\psi=i\hslash\frac{\partial\psi}{\partial t}\label{eq: Schrodinger equation}
\end{equation}
and we write

\begin{equation}
\psi(x,t)=R(x,t)e^{iS(x,t)/\hslash}
\end{equation}
where both $R(x)$ and $S(x)$ are real functions. The guidance equation
is given by

\begin{equation}
\frac{d\mathbf{X}(t)}{dt}=\frac{1}{m}\nabla S(X(t))
\end{equation}

\noindent so we can equate the fluid velocity in a hydrodynamic picture
with this. We wish to incorporate the potential from the vortex solution
into the initial state of a wave function. We assume at an initial
time $t=0$ that we have

\noindent 
\begin{equation}
\frac{S(x,0)}{m}=\phi_{f}(x)+\phi_{w}(x)
\end{equation}

\noindent where $\phi_{f}$ is the knot potential calculated above,
and $exp(im\phi_{w}(x)/\hbar)$ is single-valued, but otherwise arbitrary.
Then the initial value for the wave function is

\noindent 
\begin{equation}
\psi(x,0)=R(x,0)e^{im\left(\phi_{f}(x)+\phi_{w}(x)\right)/\hslash}\label{eq:Initial value wave function}
\end{equation}

\noindent where here $R(x,0)$ is an arbitrary positive and single-valued
function of $x$. %
{} Let us define a mapping by continuing $x$ around the filament with
a winding number of $N_{w}$ as

\noindent 
\begin{equation}
\psi(x,0,N_{w})=\psi(x,0)e^{im\Gamma N_{w}/\hslash}\label{eq: phase factor for vorticity}
\end{equation}

\noindent and we see that this is in general multi-valued due to factor
$e^{im\Gamma N_{w}/\hslash}$. Because the Schrödinger equation is
linear, this factorization will be preserved in time, so that 

\begin{equation}
\psi(x,t,N_{w})=\psi(x,t)e^{im\Gamma N_{w}/\hslash}
\end{equation}

\noindent where it is assumed here that the analytic continuation
curve is advectively adjusted to account for motion of the filament
in time so that the winding number doesn't change. And so if the wave
function is multi-valued initially, it will remain so for all time.
The vortex filament will change its shape and undulate in time as
we integrate the equations forward or backward in time, but because
the equations are equivalent to an Eulerian fluid, the Kelvin circulation
theorem will remain in effect, and the topological knot and link structure
of the filament or filaments will remain invariant. In order for the
wave function to be single-valued we must require

\begin{equation}
e^{im\Gamma N_{w}/\hslash}=1
\end{equation}

\noindent for all integer values of $N_{w}$, because the winding
number is an arbitrary integer. This then requires

\begin{equation}
e^{im\Gamma/\hslash}=1\label{eq: first single valued condition}
\end{equation}
or equivalently

\begin{equation}
m\Gamma/\hslash=2\pi K,\:for\:K\:integer\label{eq:second single valued condition}
\end{equation}
The value of $K$ determines vorticity constant and we see that a
given knot can have any value of $K$ and still be single valued,
so long as $K$ is an integer. We expect that $K$ will be a constant,
independent of time, because of the Kelvin circulation theorem, at
least ignoring the dissipative radiative effect described below. The
question is then why should this quantization condition on $K$ be
true?

\section{Requirement that the filament must be a nodal curve for the wave
function}

It is typically assumed that if a Schrödinger wave function has a
vortex filament, that it must be a nodal filament, so that the wave
function vanishes along it. The reason for this is continuity of the
wave function. In a neighborhood of a point on the vortex filament,
the wave function's phase takes on multiple values. If it does not
vanish on the filament, then the different phases in a neighborhood
would lead to a discontinuity. Since we typically assume that the
quantum mechanical wave function must be continuous, this then requires
that the complex wave function vanishes along the vortex. See for
example \citep{hirschfelder1974quantized}. If we want to create such
a state, then we must require that the initial value for $\psi$ must
vanish on the filament curve so that it will be continuous there

\begin{equation}
\psi(\mathbf{R_{f}}(\sigma),0)=0,\:\sigma\in\left[0,l\right]
\end{equation}
where $l$ is the length of the filament knot. This requires that

\begin{equation}
R(x,0)|_{x=\mathbf{R_{f}}(\sigma)}=0
\end{equation}
so we must find a single-valued function which vanishes on the filament
curve. Such a function is easy to create from positive definite integrals
of the following type

\begin{equation}
I_{n}^{f}(x)=\int_{0}^{l}\frac{d\sigma}{\left|\mathbf{x}-\mathbf{R_{f}}(\sigma)\right|^{n}},\:n\:a\:positive\:integer
\end{equation}

\noindent then, letting $D_{f}$ denote the set of points on the filament,
we have

\begin{equation}
\frac{1}{I_{n}^{f}(x)|_{x\in D_{f}}}=0,\:if\:n\geq1
\end{equation}

\noindent then we can create a wave function by writing

\begin{equation}
\psi(x,0)=\Phi(x)\frac{e^{im\phi_{f}(x)/\hslash}}{I_{n}^{f}(x)}
\end{equation}

\noindent where $\Phi(x)$ is an arbitrary smooth single-valued complex
function which falls off at infinity fast enough to make $\psi(x,0)$
normalizable. Let us also assume that $\Phi(x)$ has no nodes so that
even if raised to a non-integer power, it remains single-valued. In
order that $\psi(x,0)$ be nodal along the filament curve, we must
require the $n=2$ or greater. With these conditions met, we have
achieved a wave function which has the desired knotted vorticity,
along with the correct nodal property, and solving the Schrödinger
equation forward in time should preserve these properties. Actually
doing this time evolution requires numerical techniques in general.
If we want to study analytic solutions for vortex loops, the elegant
methods presented in \citep{bialynicki-birula2000motionof} can be
considered. These analytic methods are not as general as the methods
presented here based on the Biot-Savart law though, and they have
only been studied for the single-valued case. 

\section{Comparison with Dirac strings}

When the quantization condition that ensures single-valuedness \eqref{eq: first single valued condition}
is satisfied, the wave function with a vortex instantaneously looks
similar to the wave function for a charged particle in the presence
of a Dirac string \citep{dirac1931quantised,dirac1948thetheory,wentzel1966comments,heras2018diracquantisation},
except that there is no monopole at the end of our vortex string as
it is either a closed curve, or one that goes off to infinity in both
directions. A reasonable question to ask in this case is whether our
vortex curves are essentially Dirac strings tied in a knot. To answer
this question, we must consider some complications in the analysis
of the Dirac string.

The vector potential outside of the Dirac string can be viewed as
due to a magnetic flux inside the string, or equivalently as a chain
of magnetic dipoles. But the magnetic flux inside the string is then
mathematically subtracted out \citep{dirac1948thetheory,wentzel1966comments}
by Dirac. He developed an action principle for this situation which
required for consistency that a charged particle never pass through
the string. In the case of a wave function in the vicinity of such
a string, this would require that the wave function vanish along the
string. This is called the ``Dirac veto'', and is precisely the
nodal behavior that we have given to our multi-valued vortex solutions
in order to make the wave function be continuous. Therefore I believe
that the vortex solutions here are essentially the same as Dirac strings
when they satisfy the Dirac veto and when the single-valued quantization
condition is satisfied by our vortex.

The theory of magnetic monopoles was modified by several authors \citep{cabibbo_quantum_1962,singleton_electromagnetism_1996,wu1976diracsmonopole,zwanziger_local-lagrangian_1971}
in such a way as to eliminate the Dirac string altogether, and thus
avoid the Dirac string which was considered a problem. The physical
effect and description of a monopole should depend only on the position
of the monopole, and not on the shape of the Dirac string attached
to it. A modified Dirac string theory based on the Wu-Yang paper,
which avoided the Dirac veto was also introduced by Brandt and Primack
\citep{brandt1977avoiding}. These approaches provide a better model
of the magnetic monopole than Dirac's original version. These theories
are clearly physically different from Dirac's original theory because
of the Dirac veto. Although they are a better description of a magnetic
monopole, the original Dirac string theory, with the Dirac veto, is
very similar if not the same as our vortex model.

In this paper we go beyond the Dirac string theory, because we consider
multi-valued wave functions which were not considered by Dirac. Moreover,
although our vortex knots look just like Dirac strings when they are
single-valued, they can exist without any monopoles attached to them,
and they complement the seminal analysis of \citep{berry2001knotting,bialynicki-birula2000motionof,lloyd2017electron,bliokh2017theoryand}
of such systems.

\section{An identity for the Schrödinger equation}

The following two equations are equivalent at points of analyticity
for the real functions $R$ and S

\begin{equation}
\left[-\frac{\hslash^{2}}{2m}\triangle+V\right]Re^{iS/\hslash}=i\hslash\frac{\partial}{\partial t}\left(Re^{iS/\hslash}\right)
\end{equation}
is equivalent to

\begin{align}
\left[-\frac{\left(\nu\hslash\right)^{2}}{2m}\triangle+\left(V+\frac{\hslash^{2}}{2m}\left(\nu^{2}-1\right)\frac{\triangle R}{R}\right)\right]Re^{iS/\left(\nu\hslash\right)}\nonumber \\
=i\left(\nu\hslash\right)\frac{\partial}{\partial t}\left(Re^{iS/\left(\nu\hslash\right)}\right)\label{eq:Nonlinear equation}
\end{align}

\noindent where $\nu$ is an arbitrary complex-valued constant. Since
both sides of \ref{eq:Nonlinear equation} are analytic functions
of $\nu$, if the equation is true for real $\nu$ then it is automatically
true for complex $\nu$ by analytic continuations. Although the identity
is true for complex $\nu$, in this paper only real values of $\nu$
will be required. An elementary proof is given in \citep{davidson1979ageneralization}
where it was used as the basis for generalizing stochastic mechanics
to arbitrary values of the diffusion constant. In another paper it
was used as a symmetry for Brownian motion \citep{davidson1979adynamical}.
This identity is a kind of nonlinear gauge transformation for the
Schrödinger equation. It is true in any number of dimensions, and
therefore it can be used with multiparticle wave functions too, with
suitable rescaling of the particle coordinates to compensate for different
masses in the equation. The generalized model allows one to consider
stochastic mechanics, Bohmian mechanics, and Heisenberg operator quantum
mechanics as all part of a covering diffusion theory \citep{shucker1980davidsons,davidson1978onthe}.
In this paper we find a new application for this identity. Suppose
that $S$ is of the form of the vortex filament solution. Let us define 

\begin{equation}
\psi_{\nu}(x,t)=R(x,t)e^{iS(x,t)/\left(\nu\hslash\right)}
\end{equation}

\noindent Now consider starting at some point $\mathbf{x}$ and analytically
continuing this function along a closed curve which winds around the
vortex filament with winding number $N_{w}$ When we do this, we obtain 

\begin{equation}
\psi_{\nu}(x,t,N_{w})=R(x,t)e^{iS(x,t)/\left(\nu\hslash\right)}e^{im\Gamma N_{w}/\left(\nu\hslash\right)}
\end{equation}

\noindent we see that for certain special values of $\nu$, $\psi_{\nu}(x,t,N_{w})$
becomes single-valued, ie. independent of $N_{w}$. The condition
for this to happen is

\begin{equation}
e^{im\Gamma N_{w}/\left(\nu\hslash\right)}=1,\:\forall\:N_{w}
\end{equation}

\noindent and so, if $\Gamma\neq0$, then either $\nu=\infty$, or

\begin{equation}
m\Gamma/\left(\nu\hslash\right)=2\pi M,\:M\in\mathbb{Z}
\end{equation}

\begin{equation}
\nu=\frac{m\Gamma}{2\pi M\hslash}\label{eq:Single valued constraint equation}
\end{equation}

In order to obtain a single-valued and linear equation, we must have
$\nu=1$ which requires $\Gamma=2\pi M\hslash/m$, for some integer
value of $M$. So single-valued knotted solutions of this type can
exist in standard quantum mechanics, in the absence of any external
electromagnetic fields, provided this condition is satisfied. Note
that if the vorticity constant $\varGamma$ is zero, then $\psi_{\nu}(x,t)$
is single valued for all values of $\nu$.

When these knotted filaments are such as to produce linear as well
as single-valued wave function, then the filaments are called quantum
vortices in the physics literature. In this case, if the wave function
is continuous at the vortex, then the filament is a nodal curve for
the Schrödinger wave function as required \citep{hirschfelder1974quantized}.
In the multi-valued case, I assume that this nodal property is still
true.

For a free particle, the nonlinear term may cause radiation and relaxation
to a linear and single-valued state. But, when there is an attractive
potential present, the situation is more complicated. There might
be non-linear bound states which don't radiate in this case.

Notice that $\nu$ is not uniquely defined by \eqref{eq:Single valued constraint equation}
since $M$ can be any integer. Without loss of generality we can restrict
consideration to positive values of $M$. 

\section{Bremsstrahlung for charged particles}

For a classical charged particle undergoing acceleration, the instantaneous
radiated power is given by Larmor's formula

\begin{equation}
P_{Rad-Cl}=\frac{2}{3}\frac{q^{2}}{c^{3}}\mathbf{a}^{2}
\end{equation}

For a quantum particle, with a wave function $\psi$ which passes
through a force field producing radiation, there are a number of calculations
of Larmor's formula for scalar particles \citep{higuchi2009quantum,higuchi2006quantum,yamamoto2011firstorder,ilderton2013radiation}.
These calculations are based on rigorous second-quantized scalar electrodynamics.
Since we are considering a non-plane wave packet here, I find it more
suitable to consider a result that I obtained in \citep{davidson2004predictions}
which is very simple. It applies to the situation where a wave-packet
moves through a localized force field due to a potential $U$ that
causes radiation. This result was the lowest order approximation in
$\alpha$. The result is, for instantaneous power radiated by a non-relativistic
wave packet simply, and to a first approximation, given by the following
quantum-Larmor formula:

\begin{align}
P_{Rad-QED}(t) & =\frac{2}{3}\frac{q^{2}}{c^{3}}\left\langle \psi(t)\right|\hat{\mathbf{a}^{2}}\left|\psi(t)\right\rangle \begin{gathered}\end{gathered}
\nonumber \\
 & =\frac{2}{3}\frac{q^{2}}{c^{3}}\int d^{3}x\rho(x,t)\left(\nabla U/m\right)^{2}\label{eq:Quantum Larmor Formula}
\end{align}

My derivation of this formula in \citep{davidson2004predictions}
leaned heavily on the radiation treatment by Schiff \citep{schiff1968quantum},
which gives the most detailed description of the effects of the wave
function's form on radiation phenomenon, and not just plane wave analysis.
It is plausible then that this expression can give a reasonable approximation
to the radiation from the nonlinear equation we found in achieving
single-valuedness \eqref{eq:Nonlinear equation}. I think is worthwhile
to see if it can be reconciled with the methods of \citep{higuchi2009quantum,higuchi2006quantum,yamamoto2011firstorder,ilderton2013radiation},
but I will not attempt that here. Higher order corrections to this
formula will not change the basic feature that is important here,
and that is that the nonlinear system will not be radiation free,
but will lose energy by radiating it away. I also found that a charged
Bose-Einstein condensate beam radiates as a combination of two simple
terms \citep{davidson2004bremsstrahlung}

\begin{gather}
P_{Bose-Einstein}(t)=\nonumber \\
N^{2}\frac{2}{3}\frac{q^{2}}{c^{3}}\left\langle \psi(t)\right|\hat{\mathbf{a}}\left|\psi(t)\right\rangle ^{2}+N\frac{2}{3}\frac{q^{2}}{c^{3}}\left\langle \psi(t)\right|\hat{\mathbf{a}^{2}}\left|\psi(t)\right\rangle 
\end{gather}
where N here is the mean number of particles in the condensate. The
first term is a kind of coherent Larmor radiation term which calculates
the radiation due to a classical charge current which is proportional
to the Schrödinger probability current. In the present situation,
due to \eqref{eq:vanishing expected force for bohm force} below,
it follows that $\left\langle \psi(t)\right|\hat{\mathbf{a}}\left|\psi(t)\right\rangle =0$
for the acceleration caused by the nonlinear force term in \eqref{eq:Nonlinear equation}.
This is consistent with the fact that in general for a single-valued
free-particle Schrödinger equation, the radiation that would be generated
from the current density treated as a classical source is always zero
\citep{davidson2007quantum}. It's not clear though, that for the
multi-valued case this still holds for all higher multipole moment
terms in the radiation expansion. In any event a single particle state
radiates approximately according to \eqref{eq:Quantum Larmor Formula}.

Now let's apply the quantum-Larmor formula \eqref{eq:Quantum Larmor Formula}
to our knotted Schrödinger solutions. Let us rewrite \eqref{eq:Nonlinear equation}
as

\begin{gather}
\left[-\frac{\hslash^{2}}{2\left(m/\nu\right)}\triangle+\frac{1}{\nu}\left(V+\frac{\hslash^{2}}{2m}\left(\nu^{2}-1\right)\frac{\triangle R}{R}\right)\right]Re^{iS/\left(\nu\hslash\right)}\nonumber \\
=i\hslash\frac{\partial}{\partial t}\left(Re^{iS/\left(\nu\hslash\right)}\right)\label{eq:Nonlinear Schrodinger Equation =0000231}
\end{gather}

We can define an effective mass

\begin{equation}
m_{eff}(\nu)=m/\nu
\end{equation}

Consider a free particle case, so we set the potential $V$ to zero
in \eqref{eq:Nonlinear equation}. We really don't know how to calculate
the radiation from a multi-valued wave function. The best we can hope
for is that the usual radiation formulas can be applied only when
the wave function is single-valued. This is plausible, but definitely
not rigorously derivable from any complete theory. If experimental
evidence supports this approach, then we can test the idea further
and zero in on a better theory if needed. Recall the Bohmian quantum
mechanical potential

\begin{equation}
Q_{B}=-\frac{\hslash^{2}}{2m}\frac{\triangle R}{R}
\end{equation}

In terms of this, the nonlinear equation is

\begin{gather}
\left[-\frac{\hslash^{2}}{2m_{eff}(\nu)}\triangle-\frac{1}{\nu}\left(\nu^{2}-1\right)Q_{B}\right]Re^{iS/\left(\nu\hslash\right)}\nonumber \\
=i\hslash\frac{\partial}{\partial t}\left(Re^{iS/\left(\nu\hslash\right)}\right)\label{eq:Nonlinear Schrodinger Equation =0000232}
\end{gather}

Incidentally, it is easy to show that

\begin{equation}
\int\rho\nabla Q_{B}d^{3}x=-\frac{\hslash^{2}}{2m}\int R^{2}\nabla\left(\frac{\triangle R}{R}\right)d^{3}x=0\label{eq:vanishing expected force for bohm force}
\end{equation}

If we assume that the quantum Larmor formula \eqref{eq:Quantum Larmor Formula}
is still valid, considering this a plausible but not rigorously derivable
hypothesis, the power radiated would then be

\begin{equation}
P(t)=\frac{2}{3}\frac{q^{2}}{c^{3}}\left(\nu-\frac{1}{\nu}\right)^{2}\int\left(\frac{\nabla Q_{B}(x,t)}{m_{eff}(\nu)}\right)^{2}R(x,t)^{2}d^{3}x\label{eq:Radiated Power}
\end{equation}

or 

\begin{equation}
P(t)=\frac{2}{3}\frac{q^{2}}{c^{3}}\left(\nu^{2}-1\right)^{2}\int\left(\frac{\nabla Q_{B}(x,t)}{m}\right)^{2}\rho(x,t)d^{3}x\label{eq: Radiated power rate}
\end{equation}

This is zero if $\nu^{2}=1$, but otherwise it's positive. Now we
can see a problem since $\nu$ is not unique owing to \eqref{eq:Single valued constraint equation}.
So the question is, what determines which value of $\nu$ to use,
and what would happen to the wave function of the particle as it lost
energy due to radiation? It's difficult to say without a full theory
for radiative effects in this situation. If, in such a complete theory,
the vorticity $\varGamma$ or possibly the mass $m$ could change,
then it could change in such a way that $\nu$ approaches $1$ asymptotically.
Considering \eqref{eq:Single valued constraint equation} this implies
(assuming $M=1$ in \eqref{eq:Single valued constraint equation})

\begin{equation}
\nu\Rightarrow1\:implies\:that\:\Gamma m\Rightarrow2\pi\hslash\label{eq: single-valued limit relation}
\end{equation}
So, in this case the multi-valued solution would transiently radiate
and might approach an equilibrium state which is both single-valued
and linear. This could give a radiative explanation for the question
posed by Wallstrom \citep{wallstrom1989onthe} at least for the free
particle state. Alternatively, it's possible that the state could
just keep radiating energy away until it was so spread out in position
that the radiation rate slowly approached zero.

Let me make a mathematical argument based on continuity in favor of
the choice $M=1$ in \eqref{eq:Single valued constraint equation}.
Suppose that $\frac{m\Gamma}{2\pi\hslash}\approx1$. So the multi-valued
wave function is only slightly multi-valued. It's reasonable then
to expect, in this case, that $\nu$ should be also close to 1, and
this implies that $M=1$ in \eqref{eq:Single valued constraint equation}.
Granted this is not a compelling physical argument, but accepting
it allows us to continue and explore if there is any experimental
evidence for this phenomenon, and if there is, then a more satisfactory
resolution of this non-uniqueness problem might be found.

For the case where the particle is not free, but subject to a binding
potential $V$, then it seems plausible that some new unexpected nonlinear
solutions to \eqref{eq:Nonlinear equation} might exist that do not
radiate, and so these could be stable, and perhaps a new form of quantum
state. In this case, different values of $M$ might be interesting
to consider. One might also imagine that the vortex knots could have
some relevance for string theory.

So to sum it up, we have presented a plausibility argument that multi-valued
wave functions for free particles might well radiate, and these could
either reach a single-valued equilibrium, or else just get more and
more spread out over time, faster than free particle quantum theory
would predict, due to radiated energy loss. In the next section we
consider a system that might yield experimental confirmation of this
effect, and if so, then it would open up a new field of research into
a more complete understanding of this phenomenon.

\section{Linear superposition}

If we have two or more multi-valued solutions to Schrödinger equation,
with different values of the vorticity constant, then their suitably
normalized superposition is also a solution. However, the identity
\eqref{eq:Nonlinear equation} we used for a single vorticity constant
won't work anymore to produce a single-valued wave function in this
case. Let $\psi_{j}$ be a solution to \eqref{eq: Schrodinger equation}
with a vorticity constant $\Gamma_{j}$. Let there be $M$ such functions,
and consider their normalized superposition

\begin{equation}
\psi=\frac{1}{\mathcal{\mathcal{N}}}\sum_{j=1}^{N}\psi_{j}
\end{equation}

\noindent where $\mathcal{N}$ is a normalization factor. So $\psi$
will also be a solution to \eqref{eq: Schrodinger equation}. Now
each $\psi_{j}$ has a value of $\nu_{j}=\frac{m\Gamma_{j}}{2\pi\hslash}$
from our nonlinear identity which makes a single-valued transformed
wave function. Writing $\psi_{j}$ as $\psi_{j}=R_{j}e^{iS_{j}/\hslash}$,
we obtain a nonlinear equation for each term in the sum

\begin{equation}
\left[-\frac{\hslash^{2}}{2\left(m/\nu_{j}\right)}\triangle+\frac{1}{\nu_{j}}\left(V+\frac{\hslash^{2}}{2m}\left(\nu_{j}^{2}-1\right)\frac{\triangle R_{j}}{R_{j}}\right)\right]R_{j}e^{iS_{j}/\left(\nu_{j}\hslash\right)}=i\hslash\frac{\partial}{\partial t}\left(R_{j}e^{iS_{j}/\left(\nu_{j}\hslash\right)}\right)
\end{equation}

\noindent Here, each function $R_{j}e^{iS_{j}/\left(\nu_{j}\hslash\right)}$
is single-valued. So the linear Schrödinger equation for $\psi$ is
equivalent to a set of decoupled non-linear equations for $\psi_{j}$
whose solutions are single-valued. If the various $\psi_{j}$ are
non-overlapping, then we could argue that the radiation emitted by
them would be approximately just the quantum Larmor formula \eqref{eq:Quantum Larmor Formula}
for each one of them summed. But when the $\psi_{j}$ have overlapping
support, then it's not obvious how to estimate the amount of radiation.
However, we can see that for a free particle, the only situation which
is clearly free of radiation is if $\nu_{j}=1$ for all $j$. Thus,
even in the case of a superposition, it is quite plausible that radiation
will occur and equilibrium will be reached only when the superposition
$\psi$ becomes single-valued. If experimental evidence can be found
for this radiation, then it will provide clues on how to generalize
the standard quantum theory to describe this radiation exactly without
resorting to plausibility arguments.

\section{Some comments on the Aharonov-Bohm effect, and predicted energy loss
due to radiation}

Consider an ideal cylindrical solenoid whose centerline is the z axis
with a static magnetic flux inside it. Let it be infinitely long in
both directions. Next consider a Schrödinger wave packet of charged
particles passing around this solenoid, and assume that the wave function
vanishes at the solenoid's surface and inside it. The space is no
longer simply connected, and the phase factor is therefore not automatically
single-valued as we analytically continue the wave function around
the solenoid. This is the vector Aharonov-Bohm system. If there are
no electromagnetic fields outside the solenoid, then in any simply
connected domain that does not intersect the volume of the solenoid,
we can choose an electromagnetic phase condition that makes $\mathbf{A}=0$
in this domain, just as for the knotted vortex filament solutions
described above. The phase shift factor obtained by analytically continuing
this wave function around the solenoid once is $exp(\pm iqF_{B}/\hslash)$,
where $F_{B}$ is the magnetic flux in the solenoid and $q$ the charge
of the particle. We can determine the effective vorticity constant
by equating this phase factor with $e^{im\Gamma/\hslash}$ from \eqref{eq: phase factor for vorticity}.
We can still use the nonlinear identity \eqref{eq:Nonlinear equation}
for this case too, and so we can again transform such a wave function
into one that is single-valued but nonlinear. We again plausibly expect
radiation from this system too, given by \eqref{eq:Radiated Power}.
A toroidal solenoid can also produce the Aharonov-Bohm effect as in
the very clean electron microscope experiments in \citep{tonomura2006theaharonovbohm}.
In these experiments, the toroidal solenoid has a superconducting
cladding which requires that the magnetic flux be quantized in increments
of $2\pi\hslash c/2e$. This results in a phase shift of either 0
or $\pi$, depending on the flux in the solenoid. When the phase shift
is $\pi$, the wave function analytically continued around the solenoid
gives a minus sign making it double-valued. As the electron beam passes
around the solenoid in this case, it should lose some energy due to
radiation if the theory presented here is correct. In this case, we
expect the kinetic energy of the particle to drop as a result of radiation.
So the exiting electron energy for a $\pi$ phaseshift should be slightly
lower than for a zero phaseshift. I'm sure it would be difficult to
measure this energy drop, but perhaps it's not impossible. Probably
a better way to look for the radiation would be to detect it directly
with a lens focused on the toroidal solenoid, and imaging onto a sensitive
electromagnetic radiation detector. The $\pi$ phaseshift case should
see radiation, but the zero phaseshift should not. The signal here
would be proportional to the beam current, which would make it easier
to detect at higher currents. This seems to me to be quite a doable
experiment. There would undoubtedly be other radiation due to the
focusing lenses of the electron optics in the electron microscope
and also due to beam electrons entering the material of the solenoid,
and so a difference between the two states of the solenoid having
zero and $\pi$ phaseshift would have to be measured. Hopefully there
would be enough of a signal to show a difference.

The understanding of the time-dependent Aharonov-Bohm effect is incomplete
at this time \citep{batelaan2009theaharonovtextendashbohm,bright_aharonovbohm_2015,choudhury2019directcalculation,jing2017onthe,macdougall_stokes_2014,macdougall_revisiting_2015,singleton2013thecovariant}.
Some authors \citep{bright_aharonovbohm_2015,macdougall_revisiting_2015,macdougall_stokes_2014,singleton2013thecovariant}
have argued that under some circumstances at least a changing time-dependent
magnetic field inside the solenoid does not change the electron interference
pattern, while others seem to question this. Introducing the possibility
of transient multi-valued wave functions into the mix of possibilities
surrounding this topic might enrich it, but it also would certainly
complicate it. We might consider what would happen to the wave function
if the magnetic field inside the solenoid varied with time. We have
predicted above that the time independent Aharonov-Bohm system will
radiate. It seems that a changing magnetic field might lead to transient
multi-valued wave functions which could then radiate time-dependent
amounts of power, but this is not obvious. For time-dependent cases
that require a non-zero vector potential outside the solenoid, the
nonlinear identity can be generalized to include a vector potential
and is therefore presented in the Appendix. 

\section{Some comments on dissipation mechanisms for achieving single-valued
equilibrium}

It's not clear how to develop a theory that would take into account
the energy loss due to radiation, and the influence that this would
have on the wave function. We might try and invoke a perturbation
expansion along the lines of conventional theory. It's not obvious
that the vortex would be preserved. If it is, and if a single-valued
limit is to be reached, then \eqref{eq: single-valued limit relation}
must be satisfied. This means that either the mass or the vorticity
of the wave function must change. Particle masses are usually taken
as constants in non-relativistic quantum mechanics, although there
are variable mass modifications which I find interesting to consider
\citep{greenberger1970theoryof,greenberger1970theoryof2,greenberger1974someuseful,greenberger1974wavepackets,fock1937dieeigenzeit,stueckelberg1941lasignification,stueckelberg1941remarque,horwitz1973relativistic,horwitz2015relativistic,land1996offshell}.
This would represent another modification of standard quantum mechanics
which would allow transient states of multi-valued wave functions
to exist transiently off the mass shell, and they could stabilize
to single-valued states over time due to radiation with the mass returning
to its normal value. This phenomenon could perhaps be relevant to
the interesting radiative decay to the mass shell value for a classical
charged particle in a Stueckelberg type of off mass shell theory as
described in \citep{aharonovich2012radiationreaction}. A more prosaic
possibility would be if the mass stayed constant and the expected
value of the energy of the wave function decreased until the multi-valued
vortex either dissipated away, or became single-valued by a change
in the circulation constant $\Gamma$. Of course it's also possible
that both the mass and the vorticity constant could change. 

\section{Topological considerations}

If we ignore radiative energy loss, then the free particle Shrödinger
equation is equivalent to an inviscid compressible Eulerian fluid
described by the Madelung theory. This would be the case if the vortex
is resulting in a single-valued linear wave function, so that it definitely
doesn't radiate. Such fluids with vortex knots and links have been
the subject of much research, and besides the circulation and the
helicity, there are other topological invariants. For example, Liu
and Ricca have studied the Jones polynomial as a dynamical invariant
of an inviscid fluid \citep{liu2012thejones,liu2013tackling,ricca1998applications},
generalizing earlier results by Moffatt et al. \citep{moffatt1969thedegree,moffatt1990structure,moffatt1992helicity}.
The Jones polynomial has been a subject of much interest in quantum
field theory due to the famous paper by Witten \citep{witten1989quantum}.
In the present circumstance we have another quantum system, namely
the single particle Schrödinger equation, which can have knotted and
linked vortex tubes whose topology can be associated with an invariant
Jones polynomial. The connection with conformal quantum field theory
was made made explicit in \citep{liu2012thejones}. These various
results are directly applicable to our present theory. This seems
like a fertile area therefore for exploration. In the case of quantum
vortices, it is known that circular ring vortex loops (or unknots)
can shrink to zero and disappear \citep{bialynicki-birula2000motionof}
and the time reverse. I don't know if this same phenomenon can occur
with other knots, like the trefoil knot for example. 

One of the ingredients in the spin statistic ``theorem'' in quantum
mechanics is the single-valuedness of orbital wave functions which
is combined with topological requirements for multi-particle states
of identical particles to yield the proofs. If nature allows transient
multi-valued orbital wave functions as has been proposed here, then
it's plausible that the spin-statistics theorem might be violated
transiently also. Already it has been established that in two dimensions
anyon particles can exist \citep{Leinaas1977,wilczek_quantum_1982}
which are neither Bosons nor Fermions. Perhaps transient anyons can
exist in 3d and higher dimensional configuration space which radiatively
decay into standard Fermions and/or Bosons over time.

\section{Discussion}

A large class of multi-valued Schrödinger wave functions have been
proposed and considered here, and it has been shown, using a nonlinear
identity of Schrödinger's equation, that they can be transformed into
solutions of a single-valued but nonlinear differential equation.
These were based on knotted vortex filament fluid methods. The Biot-Savart
law commonly used in vortex analysis was supplemented by an additional
factor along the filament to create wave functions which had the necessary
nodal properties required by continuity in quantum mechanics along
such a filamentary vortex curve. This satisfies the requirement that
the wave function vanish along the vortex filament curve. Considering
Larmor type formulas for bremsstrahlung suggests that even free charged
particles with these wave functions would radiate due to the nonlinearity
of the single-valued transformed equation in this case. Only when
the wave function is single-valued, and also simultaneously satisfies
a linear Schrödinger equation will it be clearly radiation free in
the free particle case. Since radiative emission can be expected to
change the wave function, it is conceivable that after a transient
period the wave function will achieve an equilibrium state in which
it is both single-valued and linear, so that the multi-valuedness
may only be transient. This would be expected to apply also for neutral
particles which had a magnetic moment, or other electromagnetic multipole
moments, like say a neutron. This fact could be the physical origin
of the single-valued constraint that is imposed in standard quantum
mechanics. This effect would not be expected to operate for neutrinos
however, and so perhaps their wave functions are not necessarily single-valued.
If a set of non-interacting charged particles achieved a single-valued
quantum state before coming together to interact through Coulomb potentials,
then the single-valuedness would be preserved for all time because
of the properties of the Schrödinger equation. The general problem
of multi-valued solutions in an attractive potential is more complicated,
and there may be stationary or oscillatory states, and it's possible
that some of these could perhaps be stable to radiative decay. 

The theory presented here can also be applied to the vector Aharonov-Bohm
system, and in particular for time dependent solenoids this might
be interesting. The theory makes a simple prediction that electrons
would radiate electromagnetic energy which depends on the magnetic
flux in the solenoid in such systems.

These results may offer new methods of analysis for pure mathematical
knot theory as well. The time-evolution of these Schrödinger knots
could be studied in numerical simulation for example, with or without
a potential function. Scattering of vortex knots could also be considered.
The connection with the Jones polynomials is another subject for exploration.

\section*{Acknowledgements}

I wish to acknowledge Fritz Bopp for helpful correspondence. I also
wish to thank one of the reviewers for insightful suggestions.

\section*{Appendix - A generalization of the nonlinear Schrödinger identity
to include electromagnetic potentials}

The following two equations are equivalent at points of analyticity
for the real functions $R$ and S provided that the vector potential
$\mathbf{A}$ is expressed in the transverse gauge so that $\nabla\cdot\mathbf{A}=0$

\begin{equation}
\left[\frac{1}{2}\left(-i\nabla+q\mathbf{A}\right)^{2}+V\right]Re^{iS}=i\frac{\partial}{\partial t}\left(Re^{iS}\right)
\end{equation}

\begin{equation}
\begin{array}{cc}
\left[\frac{1}{2}\left(-i\nu\nabla+q\mathbf{A}\right)^{2}+\right.\\
\left.V-\frac{1}{2}\left(\frac{\left[\left(-i\nu\nabla+q\mathbf{A}\right)^{2}-\left(-i\nabla+q\mathbf{A}\right)^{2}\right]R}{R}\right)\right]Re^{iS/\nu}\\
=i\nu\frac{\partial}{\partial t}\left(Re^{iS/\nu}\right)
\end{array}
\end{equation}

The proof is elementary and straightforward but somewhat tedious.
This is true in any number of dimensions, and therefore it can be
applied to multiparticle wave equations if the coordinates of the
particles are suitably scaled to account for mass differences. It
allows the multi-valued analysis to be applied to particles in magnetic
fields

\bibliographystyle{unsrt}
\bibliography{Multi-valued_wave_function_add_refs,Multi-valued_Schrodinger_clean_2}

\end{document}